\documentstyle[12pt,epsf]{article}
\makeatletter
\long\def\@makecaption#1#2{
 \vskip 10pt 
 \setbox\@tempboxa\hbox{{\small #1: #2}}   
 \ifdim \wd\@tempboxa >\hsize
   \begin{quote}{\small #1: #2}\end{quote} 
 \else
   \hbox to\hsize{\hfil\box\@tempboxa\hfil}
 \fi}
\makeatother
\newcommand{\mdelta}{{\scriptstyle \Delta}}
\def\Journal#1#2#3#4{{#1} {\bf #2}, #3 (#4)}
\def\PRE{{\em Phys. Rev.} E}

\title{ THE EXACTLY SOLVABLE SIMPLEST MODEL FOR QUEUE DYNAMICS} 
\author{
{\sc Y{\=u}ki Sugiyama}
\thanks{e-mail address: genbey@eken.phys.nagoya-u.ac.jp} \\
{\it Division of Mathematical Science} \\
{\it City College of Mie, Tsu, Mie 514-01} \\
{\sc Hiroyasu Yamada}
\thanks{e-mail address: hyamada@allegro.phys.nagoya-u.ac.jp} \\
{\it Department of Physics} \\
{\it Nagoya University, Nagoya 464-01}
 }
\date{}

\begin{document}
\maketitle

\vspace*{-9cm}
\begin{flushright}
DPNU-96-31 \\
June 1996
\end{flushright}

\vspace*{6.5cm}

\begin{abstract}
We present an exactly solvable model for queue dynamics.
Our model is very simple but provides the essential property for such dynamics.
As an example, the model has the traveling cluster solution as well as 
the homogeneous flow solution. 
The model is the limiting case of Optimal Velocity (OV) model, 
which is proposed for the car following model to induce traffic jam 
spontaneously.
\end{abstract}

The concept of queue dynamics offers simple 1-dimensional models for 
socio-economic and complex multi-body physical systems,
such as traffic and granular transport problems. 
Recently, several approaches have been developed on this field
\cite{ns,naga,kk,ov,ks}.
We propose an exactly solvable simplest model of this kind. 
Our model describes the general aspects of queue dynamics, and it can be 
widely applicable to such systems. But the following
discussion is presented with the terminology of traffic problems.

The model is
\begin{equation}
\ddot{x}_n = a \left\{ V(\mdelta x_n) - \dot{x}_n \right\}, 
\label{eq:ovm}
\end{equation}
where
\begin{equation}
\mdelta x_n = x_{n-1} - x_{n} ,
\label{eq:hw}
\end{equation}
for each car number $n$ ($n = 1, 2, \cdots$ ). 
$x_n$ is the position of the $n$-th car, $\mdelta x_n$ is the headway of that
car. Dot denotes the time derivative. 
$a$ is a sensitivity constant, which we set
the same value for all drivers. The function $V(\mdelta x_n)$, which is called
``OV-function'', is
\begin{equation}
V(\mdelta x_n) = v_{max}\cdot\theta(\mdelta x_n - d),
\label{eq:thea}
\end{equation}
where $\theta$ is a Heaviside function. 
It decides the optimal velocity (the safety velocity)
according to the headway: (a) if the headway is less than $d$, a car should 
stop; (b) while the headway is larger than $d$, a car can accelerate 
to move with the maximum velocity $v_{max}$. 

For both cases, the movement of each car is easily derived as follows. \par
\noindent
(a) In the case of $\mdelta x_n < d$ 
\begin{equation}
x_n(t) = x_n(t_0)+{\dot{x}_n(t_0) \over a}\{1-e^{-a(t-t_0)}\},
\label{eq:s1}
\end{equation}
with the initial condition, $t=t_0$, $x(t_0)$, $\dot{x}(t_0)$.
\par \noindent
(b) In the case of $\mdelta x_n \geq d$ 
\begin{equation}
x_n(t) = x_n(t_0)+v_{max}(t-t_0)-{{v_{max}-\dot{x}_n(t_0)} \over a}
\{1-e^{-a(t-t_0)}\},
\label{eq:s2}
\end{equation}
with the same initial condition.

For the purpose of obtaining the solution of jam flow, we should consider 
the two basic processes of car moving. Suppose a jam exist in the lane, a
car moves from ``free driving region'' into jam, and another car escapes from
jam to free driving region. These two processes can be expressed by the above 
two solutions with appropriate connection conditions.
We assume the ideal case: cars stop in jam with the same distance 
$\mdelta x_J(<d)$, and move at the maximum velocity $v_{max}$ with the same 
distance $\mdelta x_F(>d)$ in free driving region. 

First, we investigate the process of car moving from jam to 
free driving region.
When the headway of the top car in a jam cluster is $d$, 
we set $t=t_0$. At this time the position of the car is set as $x_0=0$. 
The $n$-th car in the jam moves as the following formulas.

\begin{equation}
x_n(t) = - n \cdot \mdelta x_J\;\;\;\;\;\;\;\;\;\;\;\;\;\;\;(t_0 \leq t < t_n)
\;,
\label{eq:j2f-1}
\end{equation}
and
\begin{equation}
x_n(t) = - n \cdot \mdelta x_J + v_{max}\{(t-t_n)-{1 \over a}[1-e^{-a(t-t_n)}]\}\;\;\;\;\;(t_n \leq t)\;,
\label{eq:j2f-2}
\end{equation}
where $t_n$ is the time when the headway of the $n$-th car is $d$.
After $t_n$ the car begins to move and escapes from jam.
We note $t_0<t_1<\cdots<t_{n-1}<t_n<\cdots$,
and $t_n$ is defined by
\begin{equation}
\mdelta x_n(t_n) = d \;.
\label{eq:tn-j2f}
\end{equation}
From (\ref{eq:j2f-1}) and (\ref{eq:j2f-2}), the definition (\ref{eq:tn-j2f})
is written as
\begin{equation}
\mdelta x_J + v_{max}\{(t_n-t_{n-1})-{1 \over a}[1-e^{-a(t_n-t_{n-1})}]\}
= d \;.
\label{eq:dtn-j2f}
\end{equation}
We have derived the sequence of
equations for the definitions of $t_n$$(n=1,2,\cdots)$, which
has the solution for $t_n>t_{n-1}$. 
It is
easily obtained by putting $\tau=t_1-t_0=\cdots=t_n-t_{n-1}=\cdots(>0)$,
and that is the unique solution, which is given by the following equation
\begin{equation}
\mdelta x_J + v_{max}\{\tau-{1 \over a}(1-e^{-a\tau})\} = d \;.
\label{eq:tau-j2f}
\end{equation}

It is easily seen the velocity of the $n$-th car, $\dot{x}_n(t)$ of 
(\ref{eq:j2f-2}), converges to $v_{max}$ for 
sufficient large time $(t \gg t_n)$. This means the car reaches the free
driving region,
and the headway of this car becomes $\mdelta x_F$, which is expressed as
\begin{equation}
\lim_{t \rightarrow \infty} \mdelta x_n(t) = \mdelta x_F \;,
\end{equation}
with using (\ref{eq:j2f-2}).
Thus, we have obtained the following simple relation.
\begin{equation}
\mdelta x_J + v_{max}\tau = \mdelta x_F \;.
\label{eq:djtaudf}
\end{equation}

Next, we make an analogous investigation for the process of car moving from 
free driving region into jam.
When the headway of the car positioned just behind 
a jam cluster is $d$, we set $t=t_0$ 
and the position of the car $x_0=0$.
The $n$-th car in the free driving region moves as the following formulas.

\begin{equation}
x_n(t) = - n \cdot \mdelta x_F + v_{max}(t-t_0) \;\;\;\;\;\;\;\;\;\;\;\;\;\;\;
(t_0 \leq t < t_n)\;,
\label{eq:f2j-1}
\end{equation}
and
\begin{equation}
x_n(t) = - n \cdot \mdelta x_F + v_{max}\{(t_n-t_0) + 
{1 \over a}[1-e^{-a(t-t_n)}]\}\;\;\;\;\;(t_n \leq t)\;,
\label{eq:f2j-2}
\end{equation}
where $t_n$ is the time when the headway of the $n$-th car is $d$.
After $t_n$ the car begins to decelerate and moves into jam.
We note $t_0<t_1<\cdots<t_{n-1}<t_n<\cdots$,
and $t_n$ is defined by
\begin{equation}
\mdelta x_n(t_n) = d \;.
\label{eq:tn-f2j}
\end{equation}
From (\ref{eq:f2j-1}) and (\ref{eq:f2j-2}), the definition (\ref{eq:tn-f2j})
is written as
\begin{equation}
\mdelta x_F + v_{max}\{-(t_n-t_{n-1})+{1 \over a}[1-e^{-a(t_n-t_{n-1})}]\}
= d \;.
\label{eq:dtn-f2j}
\end{equation}
As the same as the previous case, we can obtain the unique solution of
the sequence of the equations for $t_n$ with the conditions, 
$\tau=t_1-t_0=\cdots=t_n-t_{n-1}=\cdots(>0)$, where $\tau$ is
given by
\begin{equation}
\mdelta x_F + v_{max}\{-\tau+{1 \over a}(1-e^{-a\tau})\} = d \;.
\label{eq:tau-f2j}
\end{equation}

The velocity of the $n$-th car, $\dot{x}_n(t)$ of 
(\ref{eq:f2j-2}), converges to $0$ for 
sufficient large time $(t \gg t_n)$, which means the car reaches 
a jam cluster and stops.
So, the headway of this car becomes $\mdelta x_J$, which is expressed  as
\begin{equation}
\lim_{t \rightarrow \infty} \mdelta x_n(t) = \mdelta x_J \;,
\end{equation}
with using (\ref{eq:f2j-2}).
Again, we have obtained the same relation as (\ref{eq:djtaudf}),
which means 
the value of $\tau$ is the same for both processes we have 
discussed above. 

Now we can solve the equations (\ref{eq:tau-j2f}), 
(\ref{eq:djtaudf})
and (\ref{eq:tau-f2j}),
thus $\mdelta x_F, \mdelta x_J$ and $\tau$ can be expressed using $a, d$ and 
$v_{max}$ as follows, 

\begin{eqnarray}
\mdelta x_F = d + {v_{max}\tau \over 2} \;, \label{eq:df}\\
\mdelta x_J = d - {v_{max}\tau \over 2} \;. \label{eq:dj}
\end{eqnarray}
$\tau$ is determined by the following equation
\begin{equation}
a\tau = 2(1-e^{-a\tau})\;,
\label{eq:tau}
\end{equation}
which has the solution $a\tau \simeq 1.59$.
As the result, each car is moving just the same manner as its front moving
car with the time delay $\tau$, which is induced in the process of obtaining
the jam flow solution. And it is proportional to 
the inverse of sensitivity $1/a$.
The collective movement of these cars forms a jam cluster.

\begin{figure}[htb]
\begin{center}
\epsfxsize = 10cm
\hfil\epsfbox{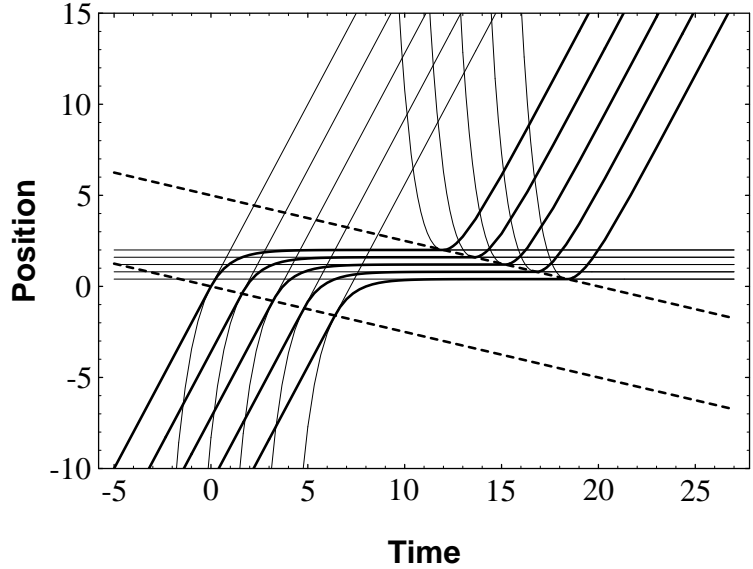}\hfill
\end{center}
\vspace{-3.7cm}
\caption{An example of jam flow. The movement of five successive cars are 
drawn by bold lines, which forms a jam cluster. 
The dashed lines show the movement of a jam
cluster, which velocity is given by (22) or (23).
\label{fig:jam}} 
\end{figure}

\begin{figure}[hbt]
\begin{center}
\epsfxsize = 10cm
\hfil\epsfbox{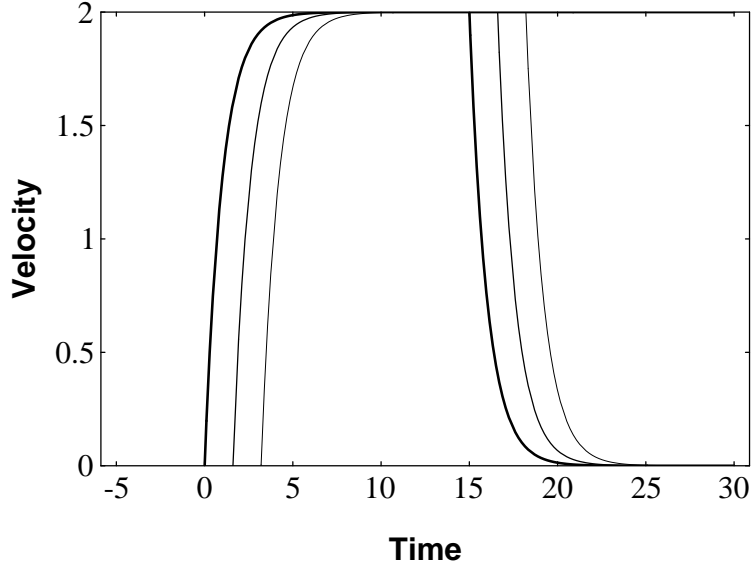}\hfill
\end{center}
\vspace{-3.7cm}
\caption{The changing of car velocity of three successive cars are drawn.
 The maximum velocity is set as $v_{max}=2$. We can see a kink like shape.
\label{fig:kink}} 
\end{figure}

In Fig.\ref{fig:jam} the moving of several successive cars are shown.
Their orbits are consisted of
(\ref{eq:f2j-1}), (\ref{eq:f2j-2}), (\ref{eq:j2f-1}) and (\ref{eq:j2f-2}) 
with the time delay $\tau$. 
Fig.\ref{fig:kink} shows the velocity of car with time 
development, which is easily obtained as the time derivatives of these 
formulas.
Rigorously say, it takes infinite time for car velocity to 
reach $v_{max}$ or $0$. Practically, 
 an appropriate finite time is enough to be considered as
infinite, as you see in Fig.\ref{fig:kink}.

\begin{figure}[htb]
\begin{center}
\epsfxsize = 10cm
\hfil\epsfbox{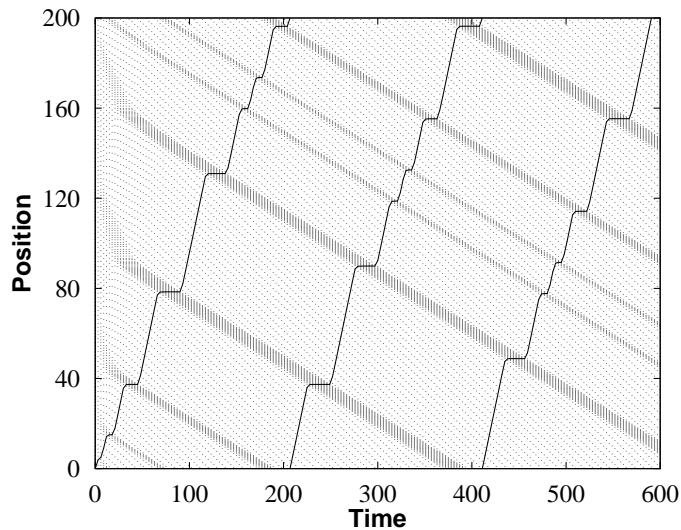}\hfill
\end{center}
\vspace{-0.5cm}
\caption{The simulation data of the positions of all cars in the circuit
with time development. The solid line shows the orbit of a sample car.
\label{sim:jam}} 
\end{figure}

In Fig.\ref{fig:jam} 
a jam cluster moves backward against the direction of car 
moving. The velocity of the cluster is defined by the moving of the front
position of the cluster, which is obtained by (\ref{eq:j2f-1}) or 
(\ref{eq:j2f-2}) as
\begin{equation}
v_{jam} = {x_{n}(t_n) - x_{n-1}(t_{n-1}) \over t_n - t_{n-1}}
        = - { \mdelta x_J \over \tau} \;.
\label{eq:vjam}
\end{equation}
The same result is given by the moving of the rear point of the cluster 
from (\ref{eq:f2j-1}) or (\ref{eq:f2j-2}) with (\ref{eq:djtaudf}).
Using the values of $\mdelta x_F$ and $\mdelta x_J$,
the velocity is also defined 
as \cite{pheno} 
\begin{equation}
v_{jam} = {{\mdelta x_{F} \cdot V(\mdelta x_{J}) - 
            \mdelta x_{J} \cdot V(\mdelta x_{F})   } 
            \over {\mdelta x_{F} - \mdelta x_{J}} }
        = - { \mdelta x_J \over \tau} \;.
\label{eq:vback}
\end{equation}
We obtain the same result as (\ref{eq:vjam}). 
We note each cluster should have the same velocity given above,
if a jam flow contains several clusters. (Actually, you can see such a case
later in Fig.\ref{sim:jam}.)
%

The profile of a jam flow is clearly described as the trajectory
in the phase space of headway and velocity $(\mdelta x, \dot{x})$. 
In order to draw this, it is enough to check the movement of two 
successive cars.
The movement is consisted of two basic processes: the moving from jam to 
free driving region and the opposite case.

First, we check the process of car moving from jam to free driving 
region.
The moving of the ($n-1$)-th and $n$-th cars 
is divided into three stages presented in Table \ref{tbl:j2f}.
%
\begin{table}[htb]
\caption{The movement of two successive cars from jam to free driving region.
 \label{tbl:j2f}}
\vspace{4pt}
\hspace{0.5cm}
\begin{tabular}{|c|p{3.7cm}|p{3.7cm}|p{3.7cm}|} \hline
$t$ & \multicolumn{3}{|p{11.1cm}|}
{$t_0$ \hspace{3.1cm} $t_{n-1}$ \hspace{2.7cm} $t_n$} 
\\ \hline\hline
$x_{n-1}$ & $\mdelta x_{n-1}<d\;:\;$ (\ref{eq:j2f-1}) &
\multicolumn{2}{p{7.4cm}|}
{\hspace{1.3cm} $\mdelta x_{n-1}>d\;:\;$ (\ref{eq:j2f-2})} \\ 
\hline
$x_n$     & \multicolumn{2}{p{7.4cm}|}
{\hspace{1.5cm} $\mdelta x_n<d\;:\;$ (\ref{eq:j2f-1})} &
$\mdelta x_n>d\;:\;$ (\ref{eq:j2f-2}) \\ 
\hline
\end{tabular}
\end{table}

\noindent
i) $t_0\leq t<t_{n-1}$ : The headway of the $n$-th car is 
$\mdelta x_J$ and its velocity is $0$. 
The $n$-th car stays the point ($\mdelta x_J$,0) in the phase space in
this period, which means the car stays in jam. 

\noindent
ii) $t_{n-1}\leq t<t_n$ : The velocity $\dot{x}_n$ is $0$. The headway is 
\begin{equation}
\mdelta x_n = \mdelta x_J + v_{max}\{(t-t_{n-1})-{1 \over a}
[1-e^{-a(t-t_{n-1})}]\} \;.
\end{equation}
From (\ref{eq:dj}) and (\ref{eq:tau}), the headway changes 
$\mdelta x_J \leq \mdelta x_n < d$ for $t_{n-1}\leq t<t_n$. Thus, 
the trajectory is the line $(\mdelta x_J,0)$ -- $(d,0)$.

\noindent
iii) $t_n \leq t < \infty $ : In this period, the headway and velocity are 
\begin{eqnarray}
\mdelta x_n &=& \mdelta x_J + v_{max}\{(t_n-t_{n-1}) 
                 +{1 \over a}[-e^{-a(t-t_n)}+e^{-a(t-t_{n-1})}]\} \;, \\
\dot{x}_n &=& v_{max}\{1-e^{-a(t-t_n)}\} \;. 
\end{eqnarray}
From these equations with using (\ref{eq:dj}) and (\ref{eq:tau}), 
we derive the following relation
\begin{equation}
\mdelta x_n = d+{\tau \dot{x}_n \over 2}\;\;\;\;\; 
(0 \leq \dot{x}_n < v_{max})\;.
\end{equation}
The trajectory is the line $(d,0)$ -- $(\mdelta x_F,v_{max})$.

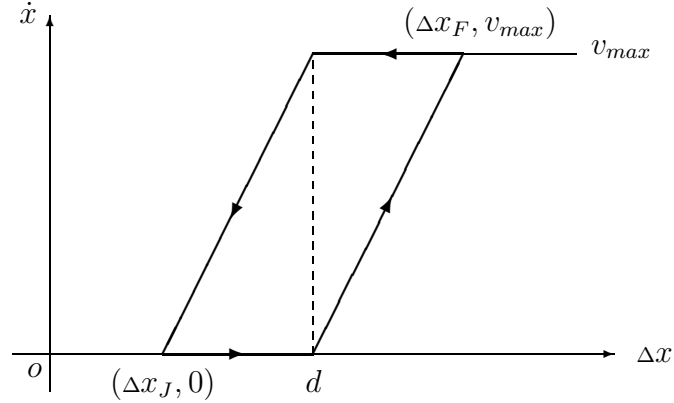
\begin{figure}[htb]
\unitlength 1mm
\begin{picture}(80,15)(-35,33)
\put(-5,0){\vector(1,0){80}}
\put(78,-1){$\mdelta x$}
\put(0,-5){\vector(0,1){50}}
\put(-4,44){$\dot{x}$}
\put(-3,-3){$o$}

\put(35,40){\line(1,0){35}}
\put(72,40){$v_{max}$}
\multiput(35,0)(0,2){20}{\line(0,1){1}}

\thicklines

\put(35,40){\line(-1,-2){20}}
\put(25,20){\vector(-1,-2){1}}
\put(8,-5){$(\mdelta x_J,0)$}

\put(35,0){\line(1,2){20}}
\put(44.5,19){\vector(1,2){1}}
\put(47,43){$(\mdelta x_F,v_{max})$}

\put(15,0){\line(1,0){20}}
\put(25,0){\vector(1,0){1}}
\put(35,40){\line(1,0){20}}
\put(45,40){\vector(-1,0){1}}
\put(34,-5){$d$}
\end{picture}
\vspace*{3.6cm}
\caption{The hysteresis loop of jam flow together with the OV-function 
$V(\mdelta x)$.
 Each car moves along this loop in
the direction of arrow with time development. 
\label{fig:hys}} 
\end{figure}

Next, we turn to the car moving from free driving region into jam.
The process is also divided into 3 stages, which is presented 
in Table \ref{tbl:f2j}.
\begin{table}[htb]
\caption{The movement of two successive cars from free driving region into jam.
\label{tbl:f2j}}
\vspace{4pt}
\hspace{0.4cm}
\begin{tabular}{|c|p{3.7cm}|p{3.7cm}|p{3.7cm}|} \hline
$t$ & \multicolumn{3}{|p{11.1cm}|}
{$t_0$ \hspace{3.1cm} $t_{n-1}$ \hspace{2.7cm} $t_n$} 
\\ \hline\hline
$x_{n-1}$ & $\mdelta x_{n-1}>d\;:\;$ (\ref{eq:f2j-1}) &
\multicolumn{2}{p{7.4cm}|}
{\hspace{1.3cm} $\mdelta x_{n-1}<d\;:\;$ (\ref{eq:f2j-2})} \\ 
\hline
$x_n$     & \multicolumn{2}{p{7.4cm}|}
{\hspace{1.5cm} $\mdelta x_n>d\;:\;$ (\ref{eq:f2j-1})} &
$\mdelta x_n<d\;:\;$ (\ref{eq:f2j-2}) \\ 
\hline
\end{tabular}
\end{table}

\noindent
i) $t_0\leq t<t_{n-1}$ : The headway of the $n$-th car is 
$\mdelta x_F$ and it's
velocity is $v_{max}$. The $n$-th car stays the point 
($\mdelta x_F$,$v_{max}$), 
which means the car moves in free driving region.

\noindent
ii) $t_{n-1}\leq t<t_n$ : The velocity $\dot{x}_n$ is $v_{max}$. 
The headway is 
\begin{equation}
\mdelta x_n = \mdelta x_F + v_{max}\{-(t-t_{n-1})+{1 \over a}
[1-e^{-a(t-t_{n-1})}]\} \;.
\end{equation}
From (\ref{eq:df}) and (\ref{eq:tau}), the headway changes
$\mdelta x_F \geq \mdelta x_n > d$ for $t_{n-1}\leq t<t_n$. 
Thus, the trajectory is the line 
$(\mdelta x_F,v_{max})$ -- $(d,v_{max})$.

\noindent
iii) $t_n \leq t < \infty $ : In this period, the headway and velocity are 
\begin{eqnarray}
\mdelta x_n &=& \mdelta x_F + v_{max}\{-(t_n-t_{n-1}) 
-{1 \over a}[-e^{-a(t-t_n)}+e^{-a(t-t_{n-1})}]\} \;, \\
\dot{x}_n &=& v_{max}e^{-a(t-t_n)} \;. 
\end{eqnarray}
From these equations with (\ref{eq:df}) and (\ref{eq:tau}), 
we derive the following relation
\begin{equation}
\mdelta x_n = \mdelta x_J+{\tau \dot{x}_n \over 2}\;\;\;\;\; 
(v_{max} \geq \dot{x}_n > 0)\;.
\end{equation}
The trajectory is the line $(d,v_{max})$ -- $(\mdelta x_J,0)$.
We summarize the above results to Fig.\ref{fig:hys}. 
The car movement in jam flow solution is represented as the square shaped 
closed loop in the phase space. All cars are moving along this loop in stable,
which can be understood as a limit cycle.

In our previous work, we proposed the car following model (OV model) \cite{ov},
whose OV function
is hyperbolic tangent instead of (\ref{eq:thea}). 
We found by the numerical simulation the model organized the limit cycle
figure called ``hysteresis loop''
, which is attractive in the whole phase space \cite{hyst}.
The difference between our simple model and the previous case 
is only the shape of hysteresis loop. 
We are convinced that the essential property for the dynamics of jam flow 
solution
of OV type model is clearly understood in our simple model.

The size of hysteresis loop $\mdelta x_F - \mdelta x_J = v_{max}\tau$,
 which determine the amplitude of jam cluster,
is characterized by the induced time delay $\tau$.
It is proportional to the inverse
of sensitivity $1/a$. The loop shrinks to
the vertical line $\mdelta x = d,\; (0< \dot{x} \leq v_{max})$ as 
$a \rightarrow \infty$, in which case jam does not appear. 

We should note the all above results do not depend on the total number of
cars and the length of lane. We do not need to set the periodic 
boundary condition on the lane. If we set them as some conditions, it is easy
to calculate the total number of cars in jam (the total sum of cluster size) 
or how long a car stays in jam.

Finally, we check the above results against the simulation data.
The aim is to know how  well the analytic solution is realized,
which we have obtained by the infinite time approximation.
We set the parameters of the model
as $a=1$, $d=2$ and $v_{max}=2$. In this case $\mdelta x_F \simeq 3.59$ and 
$\mdelta x_J \simeq 0.41$ are derived from (\ref{eq:df}) and (\ref{eq:dj}).
The simulation is performed with putting cars on the circuit, whose length
$L=200$, and the total number of cars $N=100$. 

Fig.\ref{sim:jam} is the plot of the position of all cars with time 
development. The initial condition is set as all cars are uniformly
distributed with the distance $\mdelta x=L/N=2(=d)$ and 
move with the same velocity $\dot{x}=1$. 
In this case, the initial movement is highly unstable.
We can observe the growth of jam clusters, and they are moving
backward in stable with the same velocity, which value is in agreement
with the analytic result (\ref{eq:vjam}).
\begin{figure}[hbt]
\begin{center}
\epsfxsize = 10cm
\hfil\epsfbox{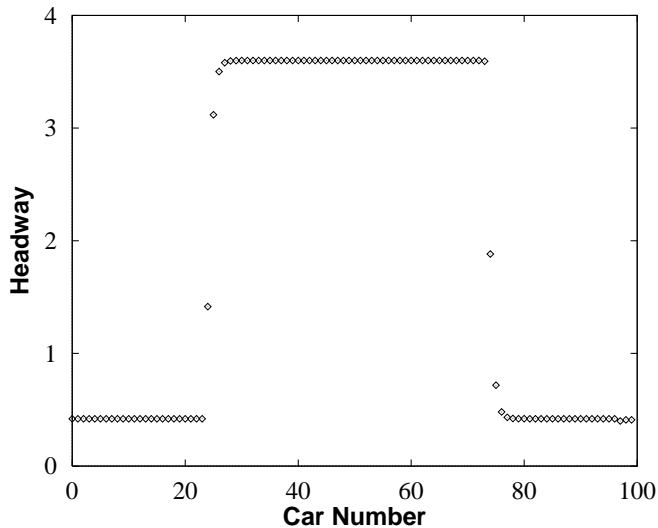}\hfill
\end{center}
\vspace{-0.5cm}
\caption{The simulation data of the distribution of headway of each car.
The data agrees with the analytic results: $\mdelta x_J \simeq 0.41$ and 
$\mdelta x_F \simeq 3.59$.
\label{sim:head}} 
\end{figure}

Fig.\ref{sim:head} is the snapshot of headway distribution of the jam flow
on the circuit. In this case cars are gathered to one big stable cluster,
which simulation is performed with
a different initial condition from the previous case. 
In both cases, the values of headway in the jam and free driving region
are just the same as analytic results.
The total number of cars in jam denoted by $N_J$ is given by the following 
formula, which does not depend on the initial conditions:
\begin{equation}
(N-N_J)\mdelta x_F + N_J \mdelta x_J \simeq L \;.
\label{eq:nj}
\end{equation}
The numerical result of $N_J$ is just the same as that of 
the analytic prediction.

\begin{figure}[htb]
\begin{center}
\epsfxsize = 10cm
\hfil\epsfbox{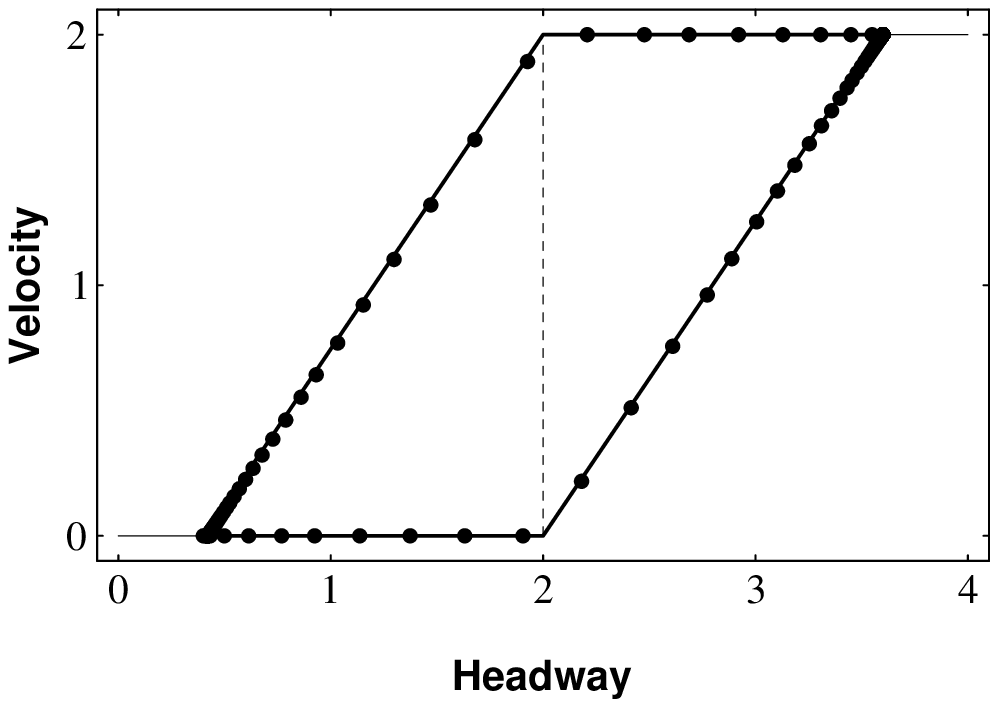}\hfill
\end{center}
\vspace{-4cm}
\caption{The simulation data of the car points of jam flow in the phase space
together with the hysteresis loop (Fig.3) and OV function.
The distribution of car points is roughly corresponding to the rapidity
of the movement of car.
\label{sim:hyst}} 
\end{figure}

Fig.\ref{sim:hyst} is the plot of car points in the phase space accumulated
over the steps after the jam cluster becomes stable.
The data points are just on the hysteresis loop, which is derived analytically.
We note the hysteresis loop is just the same for each cluster in 
both simulations. Actually, this profile is very universal, which does not
depend on initial conditions nor boundary conditions. 
This fact is easily
understood by the derivation of analytic results.
In conclusion, all simulation data shows that the jam flow is realized 
just as the analytic solution prospects.

We should discuss about the stability of the jam flow solution as well as
the homogeneous flow solutions, which can be considered as 
``free driving flow'' with no jam. 
The stable cases of the homogeneous flow are trivial. If all cars 
stop with the headway less than $d$, this situation is
stable. While, if all cars are moving with the same velocity $v_{max}$ and
with the headway larger than $d$, this flow is also stable.
In contrast to the above two cases, if the headway of a car is more than $d$ 
while the others are less than
$d$, or the opposite case, such flows are unstable and the jam flow is 
organized with
time development. And, we can observe the jam flow is stable, 
as an example in Fig.\ref{sim:jam}.

To summarize, we have proposed the simplest model for queue dynamics, which
has the travelling cluster solution (jam flow) derived analytically.
In the solution, cars are moving just the same manner with the induced time 
delay.
The amplitude of the cluster is characterized by this time delay, which is 
related to the response sensitivity. The cluster has the profile of a limit 
cycle dynamics. This shows the ``delay of changing motion'' of each car,
which means the balance of car moving in and out of a cluster,
and this microscopic balance results the formation of the macroscopic cluster.
Our simple model can provide the general understanding for queue dynamics
by comparing with another approaches \cite{nss}.

We thanks K. Nagel and M. Schreckenberg for inspiring us the idea of this work
through the discussion during the one of the author(Y.S) was staying 
in Duisburg. 
And we should comment that T. Nagatani has the work concerned with the similar 
simplest cellular automaton model based on the OV model \cite{nagapp}.




\begin{thebibliography}{99}
\bibitem{ns}K. Nagel and M. Schreckenberg, 
\Journal{\em J. Phys. {\bf I} France}{2}{2221}
{1992}; M. Schreckenberg, A. Schadschneider, K. Nagel and N. Ito, 
\Journal{\PRE}{51}{2939}{1995} e.t.c. 
\bibitem{naga}T. Nagatani, \Journal{\em J. Phys. Soc. Jpn.}{62}{3837}{1993};
\Journal{\PRE}{51}{922}{1995} e.t.c.
\bibitem{kk}B. S. Kerner and P. Konh$\ddot{\rm a}$user,
\Journal{\PRE}{48}{2335}{1993};
\Journal{\PRE}{50}{54}{1994} e.t.c.
\bibitem{hyst} M. Bando, K. Hasebe, A. Nakayama, A. Shibata and Y. Sugiyama,
\Journal{\em Japan J. of Ind. and Appl. Math.}{11}{203}{1994}; 
\bibitem{ov} M. Bando, K. Hasebe, A. Nakayama, A. Shibata and Y. Sugiyama,
\Journal{\PRE}{51}{1035}{1995}; 
\bibitem{pheno} M. Bando, K. Hasebe, A. Nakayama, K. Nakanishi, A. Shibata 
and Y. Sugiyama,
{\em J. Phys. {\bf I} France}{\bf 5}, 1389 (1995). 
\bibitem{ks}T. S. Komotsu and S.-i. Sasa, \Journal{\PRE}{52}{5574}{1995}.
\bibitem{nss}K. Nagel, M. Schreckenberg and Y. Sugiyama, in preparation.
\bibitem{nagapp}T. Nagatani, pre-print in Shizuoka Univ.
\end{thebibliography}
\end{document}